# Crustal velocity and strain rate fields in the Balearic Islands based on continuous GPS time series from the XGAIB network (2010-2013)

Alberto Sánchez-Alzola[(1)*], Carlos Sánchez[(2)], Jordi Giménez[(3)], Pedro Alfaro[(4)], Bernadí Gelabert[(5)], María J. Borque[(6,7)], Antonio J. Gil[(6,7)]

[(1)] Departamento de Estadística e Investigación Operativa, Universidad de Cádiz. Campus de la Asunción, 11405 Jerez de la Fra. (Cádiz), Spain
[(2)] Serveis d'Informació Territorial de les Illes Balears (SITIBSA), Avda. Alexandre Rosselló 13b 07002 Palma de Mallorca, Spain
[(3)] Servei d'Estudis i Planificació, D.G. Recursos Hídrics, Conselleria de Medi Ambient, Govern de les Illes Balears, Palma de Mallorca, Spain
[(4)] Departamento de Ciencias de la Tierra y del Medio Ambiente, Universidad de Alicante, Apdo. 99, 03080 Alicante, Spain
[(5)] Departament de Ciències de la Terra, Universitat de les Illes Balears, Spain
[(6)] Departamento de Ingeniería Cartográfica, Geodésica y Fotogrametría, Universidad de Jaén, Campus de las Lagunillas, 23071 Jaén, Spain
[(7)] Centro de Estudios Avanzados en Ciencias de la Tierra (CEACTierra), Universidad de Jaén, Campus de las Lagunillas, 23071 Jaén, Spain

* Corresponding author: telephone: +34 653454124, e-mail: alberto.sanchez@uca.es

**Abstract**

In this paper, we present a first estimation, using the GIPSY-OASIS software, of the crustal velocity and strain rate fields in the Balearic Islands (Spain), based on continuous GPS observations from the XGAIB network spanning the period 2010-2013. The XGAIB network consists of nine permanent, widely distributed stations that have operated continuously since 2010. In this paper, we describe the XGAIB network and the CGPS data processing and present our principle results in terms of the position time series and velocities of all of the sites, which were observed for more than three and a half years. In addition, strain tensors were estimated from the velocity field to obtain the first realistic crustal deformation model of the archipelago. The strains exhibit gradual variation across the Balearic Islands, from WNW-ESE extension in the southwest (Ibiza and Formentera) to NW-SE compression in the northeast (Menorca). These results constitute an advance in our knowledge of the tectonics of the western Mediterranean region.

**Highlights**

- We model the crustal velocity and strain rate fields in the Balearic Islands.

- We use continuous GPS observations from a permanent GNSS network.

- The strains exhibit gradual variation from WNW-ESE in the southwest to NW-SE in the northeast.

**1. Introduction**

The Balearic Islands are located on the east-northeast end of the Betic Cordillera in the western Mediterranean Sea. The westernmost part of the Mediterranean is a region of a great research interest due to its complex geodynamic setting. Several models constructed from geophysical and geologic data have been proposed for the present-day geodynamic setting of the region (Argus et al., 1989; Dewey et al., 1989; DeMets et al., 1994; Calais et al., 2003). More recently, new geodetic data have been used to improve these models (McClusky et al., 2003; Nocquet and Calais, 2003; Stich et al., 2006; Fadil et al., 2006; Serpelloni et al., 2007; Fernandes et al., 2007; Tahayt et al., 2008; Perez-Peña et al., 2010; Vernant et al., 2010; Palano et al., 2011; Koulali et al., 2011). The Balearic Islands are located in a broad plate boundary zone of slow oblique NW–SE convergence on the order of 4 to 6 mm/yr between the Nubian and Eurasian plates. While the Eurasian-Nubian plate boundary is diffuse in the region of the Iberian Peninsula (from 12ºW to 1ºE), it is concentrated in a narrow zone of the central Maghrib (Algeria) in the segment of the Balearic Islands (from 1ºE to 7ºE longitude) (Fig. 1). Therefore, the Balearic Islands appear to accommodate only a small portion of the plate convergence, which is characterised by weak seismic activity. This weak activity is the likely reason for the scarcity of studies of the active tectonics (Silva et al., 1997, 1999, 2000, 2001; Giménez and Gelabert, 2002; Giménez, 2003; Gelabert et al., 2005). Other onshore studies have described the Plio-Quaternary activity in the Balearic Promontory (Roca, 1996; Acosta et al., 2001, 2002, 2004; Maillard and Mauffret, 2013).

Previous geodetic information from the study area is limited to regional GPS studies of the western Mediterranean, focusing on the Eurasian-African plate boundary, where specific references to the Balearic Islands are scarce (Fadil et al., 2006; Serpelloni et al., 2007; Perez-Peña et al., 2010; Vernant et al., 2010; Palano et al., 2011; Koulali et al., 2011). According to Serpelloni et al. (2007), the Algerian Tell accommodates 2.7 to 3.9 mm/yr of the present-day Africa-Eurasia convergence. Those authors concluded that active shortening at rates between 1.6 and 2.7±0.6 mm/yr is occurring between northern Africa and Iberia, across the Algero-Balearic basin. They also concluded that the Africa-Eurasia convergence is not fully absorbed by the northern Africa seismic belts but is transferred to and accommodated in areas farther north.

Since 2010, an active GNNS network (XGAIB) for accurate real-time positioning and continuous GNSS data for geodetic, cartographic, and topographic work has been used in the Balearic Islands. In this paper, we present a study with the aim of modelling, for the first time, the present-day deformation in the archipelago, thus establishing a preliminary internally consistent velocity field in this area from the CGPS position time series using precise point positioning (PPP) and determining the crustal strain rates. These results were integrated with geological information to improve the understanding of the tectonics of the western Mediterranean region.

**2. Geodynamic setting**

The Balearic Islands are the emergent part of the Balearic promontory, which extends in a SW-NE direction from Ibiza to Menorca (Fig. 2). These islands have been affected by at least two tectonic phases since the late Oligocene: (1) Late Oligocene to Middle Miocene NW-SE shortening which produces NE-SW compressive structures, recognised on almost all of the islands, and (2) Late Miocene extension, which formed a set of extensional basins and produces the dismemberment of the Balearic promontory into two blocks separated by the Mallorca channel (Fig. 2).

Evidences or recent tectonic activity are very scarce in the whole promontory. Based on the sediment thickness that accumulated in the Inca Basin (Mallorca) during the Neogene, the average rate of vertical displacement of these structures between the Miocene and the present was 0.1 mm/year (Giménez and Gelabert, 2002). The evidence of recent tectonic activity in Mallorca is concentrated in the central part of the island and may be related to the NE-SW structures that interact with NW-SE minor structures. According to Sàbat et al. (2011), the Late Miocene extension is perpendicular to the NNE-SSW extensional faults which produced rollover anticlines. The aligned basins and horsts located between the Tramuntana Range and the central zone of Mallorca are interpreted as transpressive and transtensional zones associated with the sinistral movement of the main NE-SW faults of the island (Giménez and Gelabert, 2002; Giménez, 2003).

In Menorca, the Neogene extension is not as evident as it is in Mallorca. According to Gelabert et al. (2005), the southern part of the island, which is underlain nearly entirely by Upper Miocene calcarenites, is dominated by a broad anticline with a NNE-SSW-trending axis, and this structure was related to inversion tectonics of a NNE-SSW Neogene normal fault.

In the Pitiusas, the post-Alpine structures are even less evident due to the low tectonic activity. In Ibiza, the compressive structures are cut by NNW-SSE faults. In Formentera, where the oldest rocks are Late Miocene, no evidence of recent activity has been observed, although the morphology of the island is controlled by NE-SW and NW-SE fractures.

Several investigations (Roca, 1996; Acosta et al., 2001, 2002 and 2004; Maillard and Mauffret, 2013) describe NE-SW and NW-SE normal and strike-slip faults affecting the Plio-Quaternary materials in the offshore area of the Balearic Islands.

The distribution of seismicity in the Balearic Islands provides additional information regarding the recent tectonic activity. Much of the recorded seismicity is located in the central part of the promontory (Mallorca and the Mallorca channel), characterized by low to moderate magnitude earthquakes associated with NE-SW faults. In Mallorca, recorded and felt historic seismicity has been concentrated in the central part of the island and can be associated with the Neogene NE-SW faults (Fig. 2). The seismicity in Menorca may be associated with the primary ENE-WSW Neogene structure that separates the Migjorn and Tramuntana regions. No significant earthquakes have been recorded in the Pitiusas Islands; the nearest seismic activity is located in Mallorca and the Eivissa channels (Fig. 2).

Unfortunately, no focal mechanisms in the islands are available. The nearest available focal mechanism was estimated for an earthquake located approximately 100 km north of Menorca, and it is consistent with a NE-SW or NW-SE vertical structure exhibiting purely strike-slip motion (Braunmiller et al., 2000) (Fig. 1). In contrast, the focal mechanisms of earthquakes located between the Balearic Islands and the south-east coast of the Iberian Peninsula (Alicante province) and those located in the south of the Algerian Basin are primarily associated with E-W reverse faults, and earthquakes in the area of the Gulf of Valencia are mainly associated with N-S normal faults. Most of these focal mechanisms are consistent with regional N-S compression and orthogonal extension. The orientation of the structures determines the nature of the faulting: reverse motion on the E-W-trending faults, strike-slip motion on the NE-SW and NW-SE faults, and normal motion on the N-S faults (Fig. 1).

## 3. CGPS data and processing

There has been a significant increase in the number of regions and autonomous communities in Spain that count on active GNSS networks for accurate real-time positioning and continuous GNSS data for geodetic, cartographic, and topographic work. Two-thirds of Spain's autonomous communities currently have a GNSS network offering continuous data for real-time positioning and post processing purposes. The Balearic Islands is one of Spain's regions that possesses one of the newest networks, which has provided GNSS data since 2010.

The precise point positioning (PPP) method (Zumberge et al., 1997), using zero-difference GPS observables, has become a valuable and reliable tool for investigating various geophysical processes at the millimetre level (Larson et al., 2004; Smith et al., 2004, Kouba, 2005; Hreinsdóttir et al., 2006). This technique has been used recently to estimate the velocity fields (Pérez et al., 2003, Teferle et al., 2007); specifically, the results from periodic investigations, continuous GPS observations and differential processing have been compared and validated.

*3.1.* *XGAIB network*

The XGAIB network (Xarxa de Geodèsia Activa de les Illes Balears) is an active geodetic network belonging to the Serveis d'Informació Territorial de les Illes Balears SITIBSA (http://xarxagnss.caib.es). This network consists of nine GNSS stations, which have been operating since 2010. The primary objective of this network is to provide single station-based differential corrections (single-base RTK) for DGPS and RTK positioning and provide RINEX data to users for geodetic, cartographic, and topographic studies.

The XGAIB stations are evenly distributed across the islands of the archipelago. There are five stations in Mallorca ("MALL" (Palma de Mallorca), "TRAM" (Tramuntana), "SINE" (Sineu), "JORD" (Colonia Sant Jordi), and "BONA" (Cala Bona)); two in Menorca ("MENC" (Ciutadella) and "ALOR" (Alaior)); one in Ibiza ("EIVI") and one on the island of Formentera ("FORM"). The MALL station (Mallorca) has been operated by the EUREF Permanent Network (EPN) since 2000. All of the stations have LEICA GRX1200+GNSS receivers and LEIAR25 choke ring antennas, except for MALL, which has a LEICA GRX1200GGPRO receiver and a LEIAT504 choke ring antenna. The XGAIB network uses the coordinates of the Spanish National Reference System

ETRS89, developed by the Spanish National Geographic Institute (IGN). It is important to note that the antennas have not been changed since the network began operating. Fig. 3 shows the distribution of the XGAIB and EUREF stations across the Balearic Islands.

The Balearic geodetic are located on geological formations of various ages and rock types. The FORM, MENC, and MALL stations were constructed on Upper Miocene carbonate materials, the EIVI station was constructed on top of a hill underlain by Quaternary material, the TRAM station was constructed on Triassic shales, the SINE station was established on Burdigalian marls, the BONA and JORD stations were built on Pliocene-Quaternary calcarenites and the ALOR station was constructed on Jurassic dolomites.

### *3.2. CGPS position time series*

We used version 6.2 of the GIPSY-OASIS software developed by the JPL in March 2013. This program has a basic gd2p.pl module that works with single receiver data giving a static point positioning in a standard stacov coordinate file. Using certain basic flags joined to the gd2p.pl module (Bertiger et al., 2010), the software takes the 30-second 24-hour RINEX file from the XGAIB network and processes the data downloading all of the files needed for the PPP processing from the JPL server.

One of the primary advantages of using this software is the single coordinate system on which the files are based. From the last reprocessing of products in 2011, all of the position results are given in the IGS08 coordinate system, and the time series is thus free of any offsets or jumps due to changes in the reference frame. Furthermore, it is possible to increase the accuracy of the coordinate calculus using a higher rate of sampling of the clock file (as short as 5 seconds), which is an advantage in short time series or kinematic studies.

Investigations of geophysical processes, such as plate tectonics and deformation, require accurate long-term estimates of station motion, such as its velocity, with realistic error bounds. The XGAIB network provides three and a half years (March 2010 to September 2013) of 1-second sampling interval RINEX files, which were subsequently re-sampled to develop 30-second files for processing using GIPSY-OASIS. The data from stations belonging to EPN were downloaded from an official FTP repository in their original 30-second sampling format.

Focus on the methodology and internal considerations of processing, a standard GIPSY-OASIS procedure with the gd2p.pl module was applied to all of the data sets to produce the position time series for the period 2010-2013. First, GIPSY-OASIS downloaded JPL final ephemeris and pole products from the JPL FTP repository (ftp://sideshow.jpl.nasa.gov/pub/JPL_GPS_Products/Final) in a single IGS08 coordinate system. To fix ambiguities with zero difference in the PPP processing, the strategy described in Bertiger et al. (2010) using the wide lane phase bias file (wlpb) available on the JPL was applied. Furthemore, The GMF troposphere mapping function (Boehm et al., 2006) and ionosphere TEC values was also used. In addition, the FES2004 ocean tide loading model (Lyard et al., 2006) from the Onsala Space Observatory and WahrK1 terrestrial tide model (Wahr, 1985) were considered. We modelled the hydrostatic component of the zenith tropospheric delay and estimated the wet component using the appropriate flag in the gd2p.pl module. The IGS08_week.ATX antenna calibration file was applied to correct the Antenna Phase Centre. An elevation mask of 10° was set up as well. A 300-second sampling of the RINEX observations was used for the coordinate estimation. Eventually, no network adjustment was applied. Only the daily PPP solutions were selected, and no dependencies between stations were thus considered.

Fig. 4 shows the de-trended position time series of the XGAIB network stations in terms of their horizontal components, east and north. When analysing the dispersion of the time series, we observed signals with predominantly annual and semi-annual periods in both components with a similar effect.

### 3.3. CGPS-derived velocity field

The precise coordinates were converted from their geocentric XYZ coordinates into their east, north, and vertical components. Prior the estimation of the velocity field, certain outliers were eliminated from the time series using a specific threshold based on the deviation of each station. To obtain the absolute velocity field of the Balearic Islands, the time series were included in CATS GPS coordinate time series analysis software, which uses a maximum likelihood estimation to fit a multi-parameter model (Williams, 2008). To develop the best adjustment, periodicity parameters were also considered. Annual, semi-annual, and 13.66-day periods (Penna and Stewart, 2003) dominate the daily series (Teferle et al., 2007).

To highlight the differences between the absolute velocities and the general movement of the Eurasian plate, the residual velocity field was obtained. We used GEODVEL plate tectonic model (Argus et al., 2010) with no net rotation (NNR) condition and the Eurasian plate as reference. For this purpose we used the appropriate transformation parameters of this model with the geographical coordinates of our stations as inputs. After that, we subtracted the velocity of the model to our estimated absolute velocities obtained via CATS. Table 1 shows the continuous GPS-derived absolute velocities and associated errors in the IGS08 reference frame. The accuracy of the estimation was obtained using the weighted least squares method (Koch, 1999). Fig. 5 shows the residual velocity field with respect to the GEODVEL model (Argus et al., 2010) and 95% confidence ellipses.

*3.4. Strain rates*

Based on the absolute velocity field previously computed, we calculated the strain tensors and the dilatation and extension with the maximum shear strains (Voosoghi et al., 1976; Grafarend and Voosoghi, 2003; Pietrantonio and Riguzzi, 2004). The software module in Matlab, i.e., GRID_STRAIN, was used (Teza et al., 2008). Fig. 6 shows the principal axes of the strain tensors, showing only significant grid points after performing the strain calculation. Fig. 7 shows the contour map of change in area; red denotes areas of contraction, and yellow denotes areas of extension.

**4. Discussion and conclusions**

The absolute velocity trend in the Balearic Islands is to the northeast, similar to the regional velocity in this part of the western Mediterranean. The calculated absolute velocity field consists of components of 19.85-21.27 mm/yr to the east and 16.00-16.86 mm/yr to the north, with averages of 20.45 mm/yr to the east and 16.45 mm/yr to the north. The network displays an internally consistent velocity overall, with only minor deviations in direction and magnitude at a few sites. The estimation errors are consistent among the stations, with mean values of approximately 0.05 mm/yr E and 0.05 mm/yr N (Table 1).

Fig. 5 shows the residual velocity field in the Balearic Islands, i.e., the values with respect to the GEODVEL model and Eurasian plate as reference. These residual velocity components are less than 0.83 mm/yr E and 1.49 mm/yr N. There is good correspondence between the residual velocity vectors of the sites in Menorca and the BONA station in Mallorca, whereas the EIVI and FORM residual velocities display certain differences with respect to those of the model. However, the vectors of the western Mallorca sites display roughly similar directions and small magnitudes.

An analysis of the strain rates (Figs. 6 and 7) yielded the following results: (1) a greater NW-SE shortening in Menorca and eastern Mallorca (BONA site), (2) E-W extension in central Mallorca (near the MALL and TRAM sites), (3) low rates of deformation in southeast Mallorca (JORD site), and (4) WNW-ESE extension in Formentera and Ibiza. In general, a gradual variation from WNW-ESE extension in the southwest islands (Pitiusas) to NW-SE shortening in the northeast (Menorca) was observed across the Balearic Islands.

These estimated strain rates agree with the geologic data. In Menorca, the broad NNE-SSW Migjorn anticline (MA Fig. 8) (Gelabert et al., 2005), which also affects the Tramuntana area of Menorca, is consistent with the NW-SE shortening inferred from the geodetic data. In Mallorca, the rollover anticlines of Llucmajor and Santa Margalida, which are related to the NNE-SSW-trending Sa Pobla and Palma normal faults, are compatible with the E-W extension inferred from the GPS data. Finally, the WNW-ESE extension observed in Ibiza and Formentera can be attributed to the NE-SW-trending normal faults and the NNW-SEE faults that control the structure of these two islands and their nearby submarine areas (Fig. 8).

ACCEPTED MANUSCRIPTBased on the velocity vectors, it is possible to assign the nine GPS sites to 6 blocks separated by faults: (1) the island of Menorca, (2) the Tramuntana Range of Mallorca (TRAM site), (3) central Mallorca, with its Neogene basins and Central Ranges, (4) the Llevant Ranges of Mallorca (BONA site), (5) the island of Ibiza, and (6) the island of Formentera (Fig. 8). The three Mallorca island blocks coincide with the primary ranges of the island and are bounded by NNE-SSW and NE-SW Neogene faults. The Mallorca blocks are separated from the Menorca block by an inferred NW-SE Menorca channel fault and from the Ibiza and Formentera blocks by a set of NW-SE faults located in the Mallorca channel, as described by Acosta et al 2002 (Fig. 8). We infer that these 6 blocks are controlled by a regional strike-slip regime with N-S shortening and E-W extension and are separated by NNE-SSW to NE-SW sinistral-normal faults and NW-SE dextral-normal faults.

The eastward velocity component of the Menorca block (block 1) is the largest, and the northward component of the Menorca block is the smallest. These values are consistent with the dextral movement on the inferred NW-SE Menorca channel fault (Fig. 8). The larger eastward component velocity of the Central Ranges block (block 3) with respect to the Tramuntana Range block (block 2) in Mallorca is consistent with the sinistral movement on the NE-SW faults bounding the Tramuntana and the Central Ranges (Sencelles fault). The three defined blocks in Mallorca also have larger eastward component velocities than does the Ibiza block (block 5), which is most likely attributable to the dextral movement on the NW-SE faults of the Mallorca channel. Extensional movement on the nearly N-S faults or sinistral movement on the NE-SW faults located between Ibiza and Formentera is consistent with a larger eastward velocity vector of the Formentera block with respect to the Ibiza block. The differences

Page 11 of 33

between the northward velocity components of the blocks are significantly less than are those of the eastward components. The larger northward component of the Llevant Ranges block with respect to the Central Ranges in the Mallorca block is most likely related to the sinistral movement on the NNE-SSW fault bounding the Llevant Ranges.

It should be noted that although the strain rate directions are consistent with the geologic data, there is a discrepancy between the magnitude of the estimated deformation rates and the seismic activity in the island group. The highest seismicity is located in Mallorca and the Mallorca channel, where the lowest estimated strain rates were observed.

This first quantification of the present-day crustal deformation of the Balearic Islands integrated with geological information will be useful to regional geodynamic studies of the western Mediterranean. The velocity field and strain rates are generally consistent with the distribution of geologic structures. Certain small discrepancies may be attributed to the short time span of the geodetic data in an area where the expected residual velocities are less than 1 mm/yr (generally less than 0.5 mm/yr). Future work integrating the geodetic data, new onshore and offshore geological and geophysical studies of the Balearic Promontory, and new seismologic information provided by future earthquakes will help develop a better understanding of the present day geodynamic setting.

**Acknowledgements:** This research was partly funded by the Spanish Ministry of Science and Innovation through the AYA2010-15501, CGL2011-30153-C02-02 and CSD2006-0041 projects (European Regional Development Fund-ERDF). The authors thank the Serveis d'Informació Territorial de les Illes Balears (SITIBSA) for the XGAIB network GPS data and the two anonymous referees in the revision stage of the paper for their constructive comments. A few of the figures were generated using Generic Map Tool (GMT) (Wessel and Smith, 1998) and MGDS Global Multi-Resolution Topography (web map server: http://www.marine-geo.org/tools/web_services.php).

ACCEPTED MANUSCRIPTGPS evidence for roll back of a delaminated subcontinental lithospheric slab beneath the Rif Mountains. Morocco. Geology 34, 529-532.

Fernandes, R., Miranda, J., Neijninger, B., Bos, M., Noomen, R., Bastos, L., Ambrosius, B., Riva, R., 2007. Surface velocity field of the Ibero Maghrebian segment of the Eurasia Nubia plate boundary. Geophysical Journal International 169, 315-324.

Fornós, J.J., Gelabert, B., Ginés, A., Ginés, J., Tuccimei, P., Vesica, P., 2002. Phreatic overgrowths on speleothems: A useful tool in structural geology in littoral karstic landscapes. The example of eastern Mallorca (Balearic Islands). Geodinamica Acta 15, 113–125.

Gelabert, B., Fornós, J.J., Pardo, J.B., Rosselló, V.M., Segura, F., 2005. Structurally controlled drainage basin development in the south of Menorca (western Mediterranean, Spain). Geomorphology 65, 139-155.

Giménez, J., 2003. Nuevos datos sobre la actividad post-Neógena en la Isla de Mallorca. Geogaceta 33, 79-82.

Giménez, J., Gelabert, B., 2002. Análisis de la actividad tectónica reciente en la isla de Mallorca. III Asamblea Hispano Portuguesa de Geodesia y Geofísica.

Grafarend, E.W., Voosoghi, B., 2003. Intrinsic deformation analysis of the Earth's surface based on displacement fields derived from space geodetic measurements. Case studies: Present-day deformation patterns of Europe and of the Mediterranean area (ITRF data sets). Journal of Geodesy 77, 303-326.

Hreinsdóttir, S., Freymueller, J.T., Bürgmann, R., Mitchell, J., 2006. Coseismic deformation of the 2002 Denali fault earthquake: Insights from GPS measurements. Journal of Geophysical Research 111.

Koch, K.R., 1999. Parameter Estimation and Hypothesis Testing in Linear Models.
17

Page 11 of 33

ACCEPTED MANUSCRIPT19Page 11 of 33

**FIGURE CAPTIONS**

Fig. 1. Seismicity in the western Mediterranean region during the period 1998-2013, with earthquakes of M>3.0 (ESMC Catalogue http://www.emsc-csem.org/Bulletin/search.php?filter=yes). Focal mechanisms are from the following sources: IGN database (http://www.ign.es/ign/layoutIn/sismoPrincipalTensorZonaAnio.do); Global CMT Catalog search (http://www.globalcmt.org/CMTsearch.html) and Buforn et al., 1995.

Fig 2. Main post-alpine structures around the Balearic Islands. Seismicity recorded by the IGN (between -0.5ºW and 6ºE) and by the EMSC (between 6ºE and 8ºE) and felt epicentres (MSK scale) are also shown (modified from the IGN Catalogue).

Fig 3. XGAIB and EUREF site locations on the Balearic Islands.

Fig. 4. De-trended position time series of the XGAIB network stations computed with GIPSY-OASIS software (north and east components in meters).

Fig. 5. Residual velocity field with respect to GEODVEL tectonic model with NNR condition and 95% confidence ellipses. Reference Eurasian plate.

Fig. 6. Map of the principal axes of the 2D strain rate tensor in units of $10^{-8}$ strain/yr. Blue and red colours show extension and contraction respectively.

Fig. 7. Contour map of change in area (strain/yr). Yellow and red colours show extension and contraction respectively.

Fig. 8. Geodynamic interpretation of the obtained GPS velocities. Main post alpine faults and basins are indicated (PF: Palma fault; SF: Sencelles fault; SJF: Sant Joan fault; CF: Campos fault; LA; Llucmajor anticline; SMA: Santa Margalida anticline; PB: Palma basin; IB: Inca basin; SPB: Sa Pobla basin; CB: Campos basin).

Supplementary data (e-component). CGPS-derived absolute horizontal velocity field in IGS08 frame and 95% confidence ellipses.

**TABLE CAPTIONS**

Table 1. CGPS-derived absolute velocities and associated errors in IGS08 reference frame. Residual velocities with respect to GEODVEL model. Units are in mm/yr.

|  | Coordinates | | Absolute velocities (mm/yr) | | | | GEODVEL (mm/yr) | | Residual velocities (mm/yr) | |
| --- | --- | --- | --- | --- | --- | --- | --- | --- | --- | --- |
| Site | Lat. (°N) | Long. (°E) | VE | σE | VN | σN | VE | VN | VE | VN |
| ALOR | 39.93436 | 4.14015 | 20.63 | ±0.05 | 16.26 | ±0.04 | 20.50 | 15.31 | 0.13 | 0.95 |
| BONA | 39.61369 | 3.39227 | 20.20 | ±0.05 | 16.86 | ±0.05 | 20.44 | 15.37 | -0.24 | 1.49 |
| EIVI | 38.95119 | 1.40687 | 19.85 | ±0.05 | 16.46 | ±0.05 | 20.23 | 15.51 | -0.38 | 0.95 |
| FORM | 38.70528 | 1.42879 | 20.58 | ±0.04 | 16.67 | ±0.05 | 20.28 | 15.51 | 0.30 | 1.16 |
| JORD | 39.31494 | 2.99817 | 20.61 | ±0.05 | 16.38 | ±0.06 | 20.43 | 15.40 | 0.18 | 0.98 |
| MALL | 39.55262 | 2.62455 | 20.37 | ±0.05 | 16.57 | ±0.05 | 20.32 | 15.42 | 0.05 | 1.15 |
| MENC | 40.00061 | 3.83123 | 21.27 | ±0.05 | 16.00 | ±0.05 | 20.44 | 15.33 | 0.83 | 0.67 |
| SINE | 39.64590 | 3.01535 | 20.51 | ±0.05 | 16.41 | ±0.05 | 20.37 | 15.40 | 0.14 | 1.01 |
| TRAM | 39.81844 | 2.89161 | 20.05 | ±0.07 | 16.40 | ±0.08 | 20.31 | 15.41 | -0.26 | 0.99 |

Figure

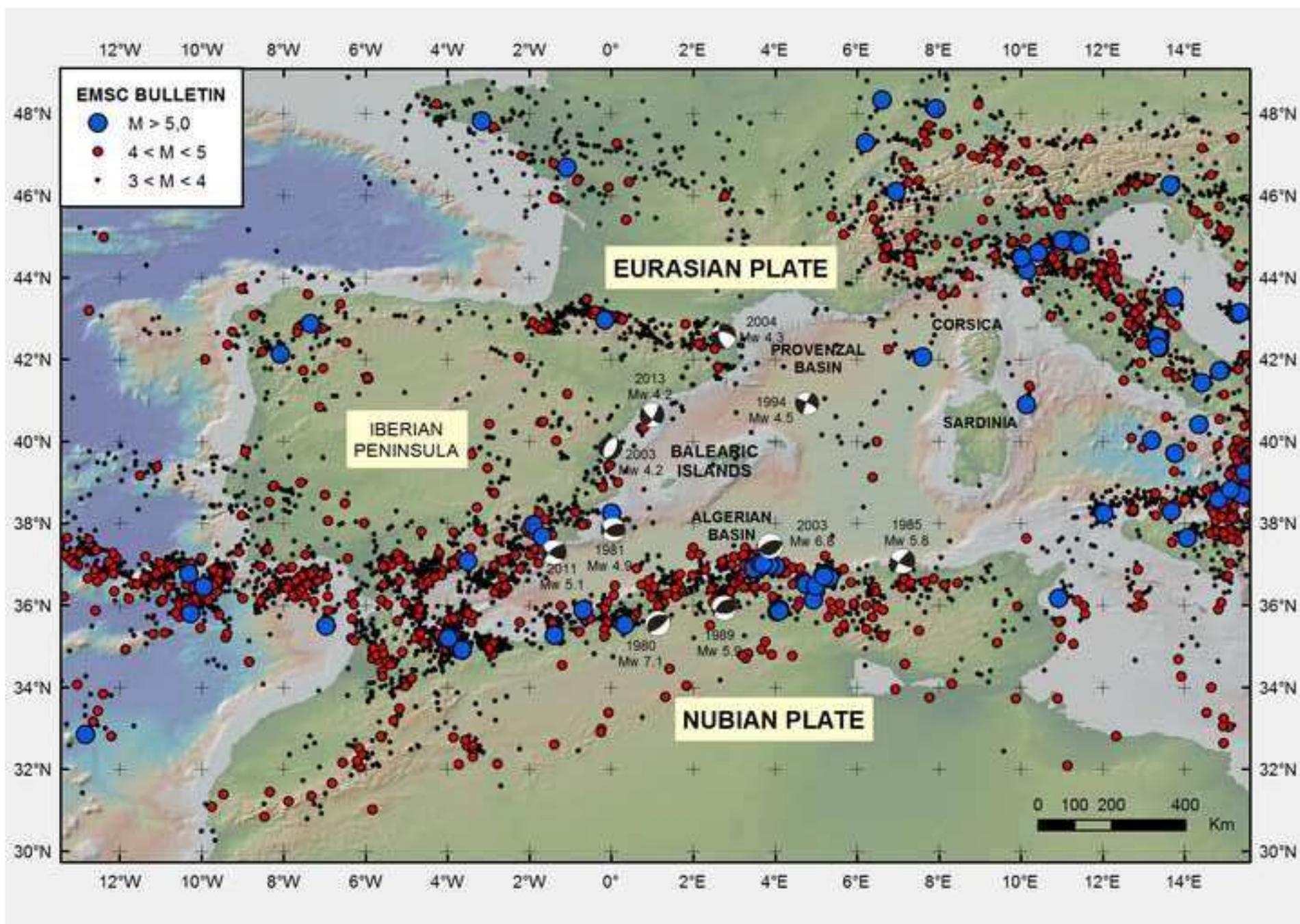

Figure

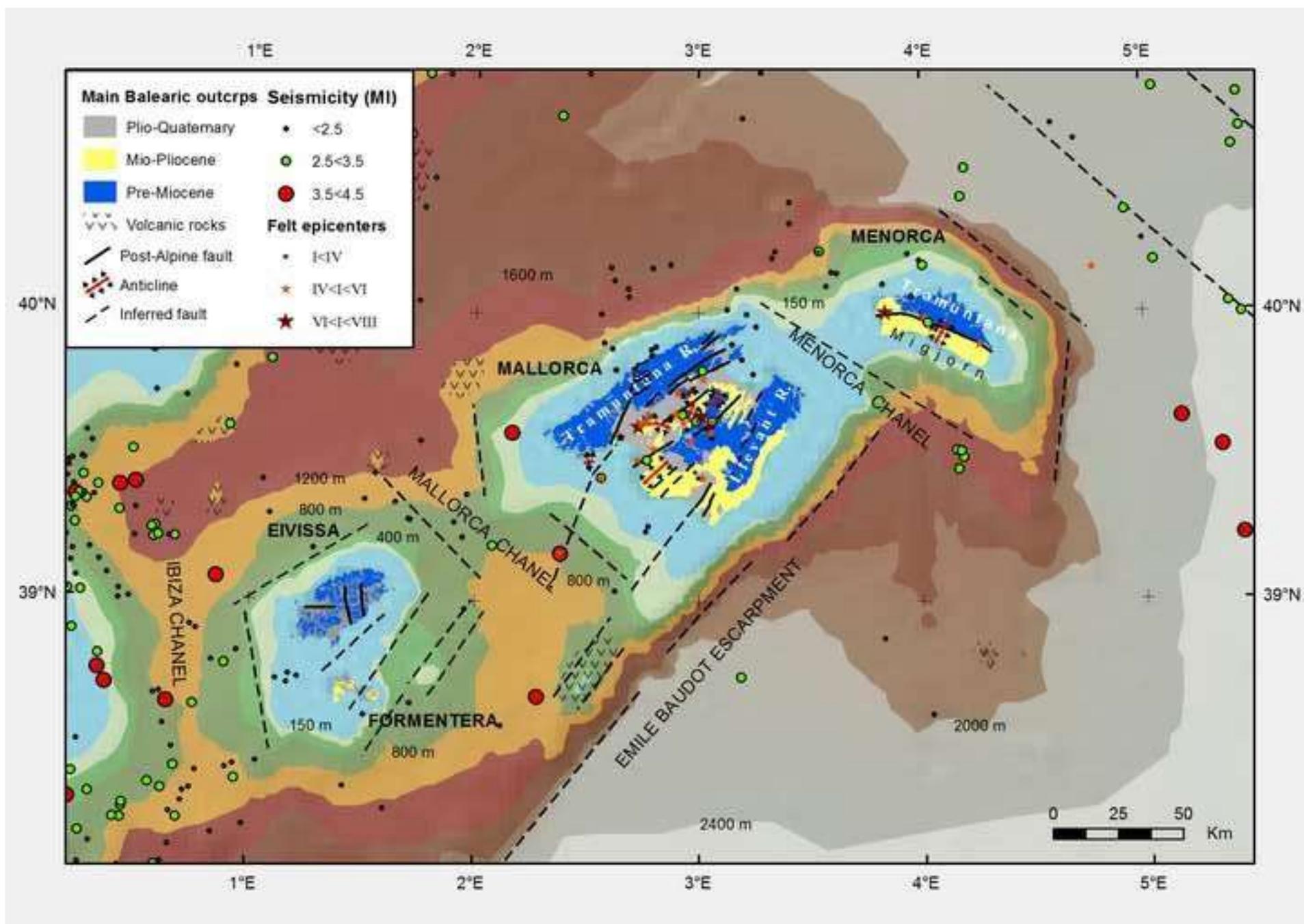

**Figure**

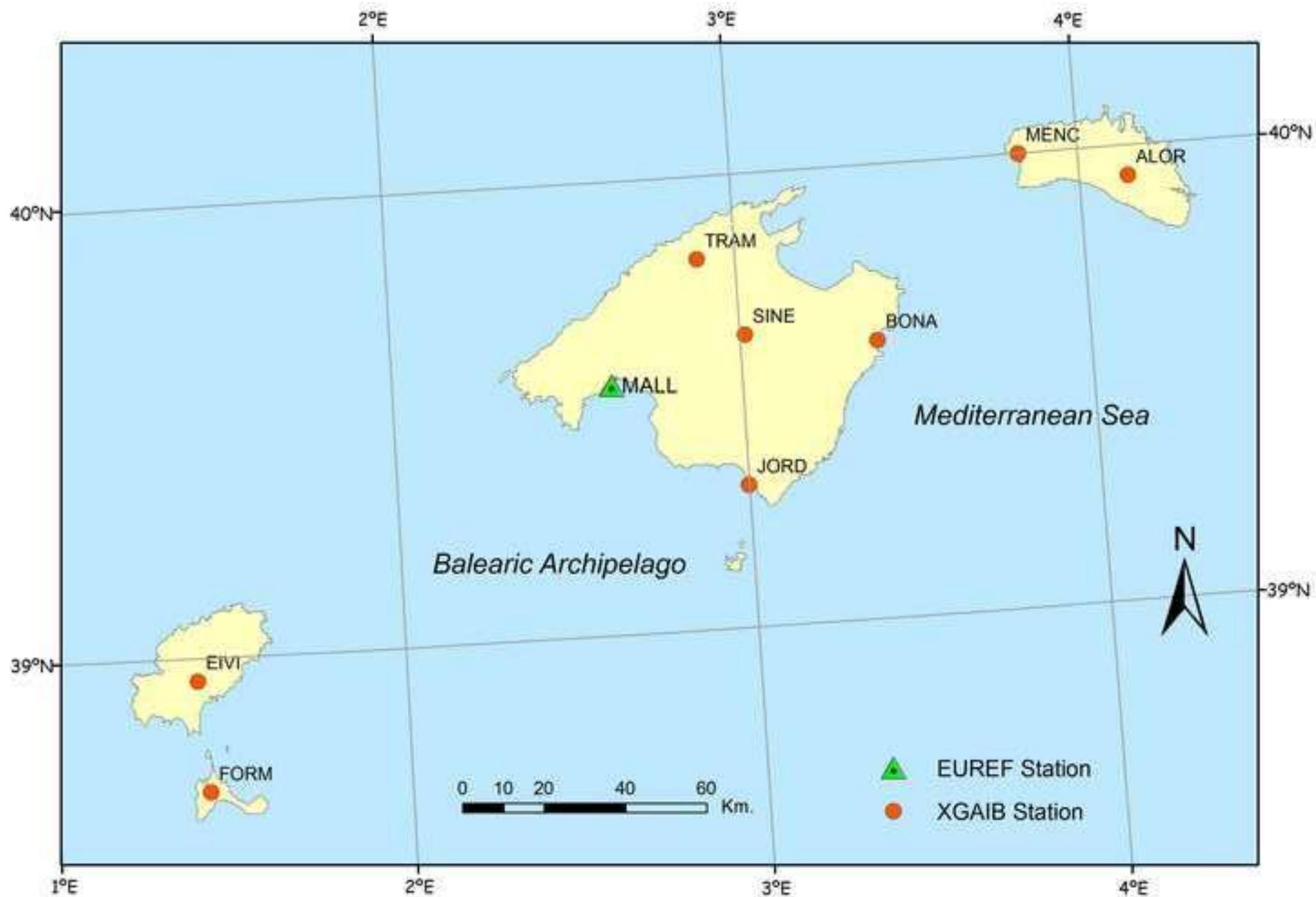

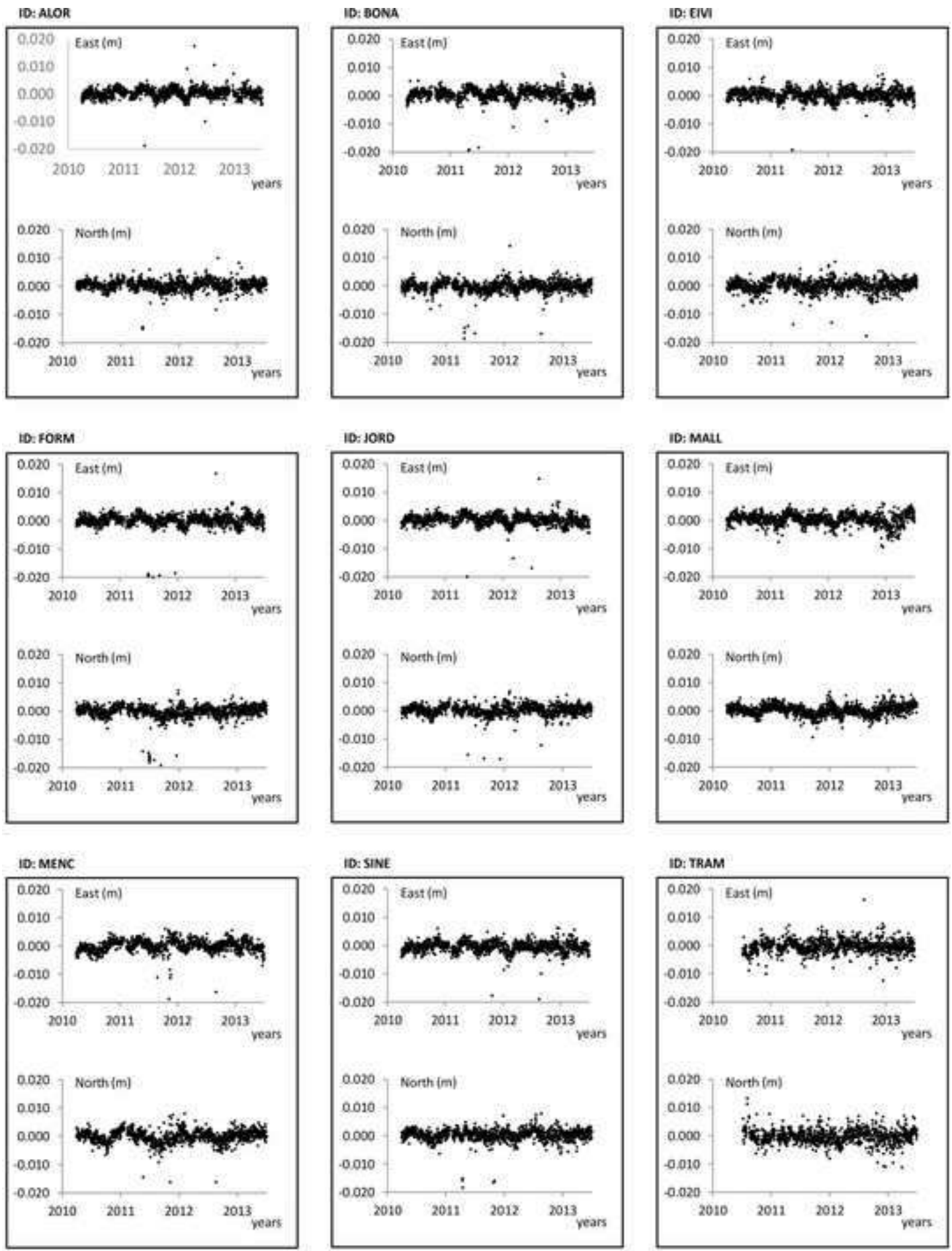

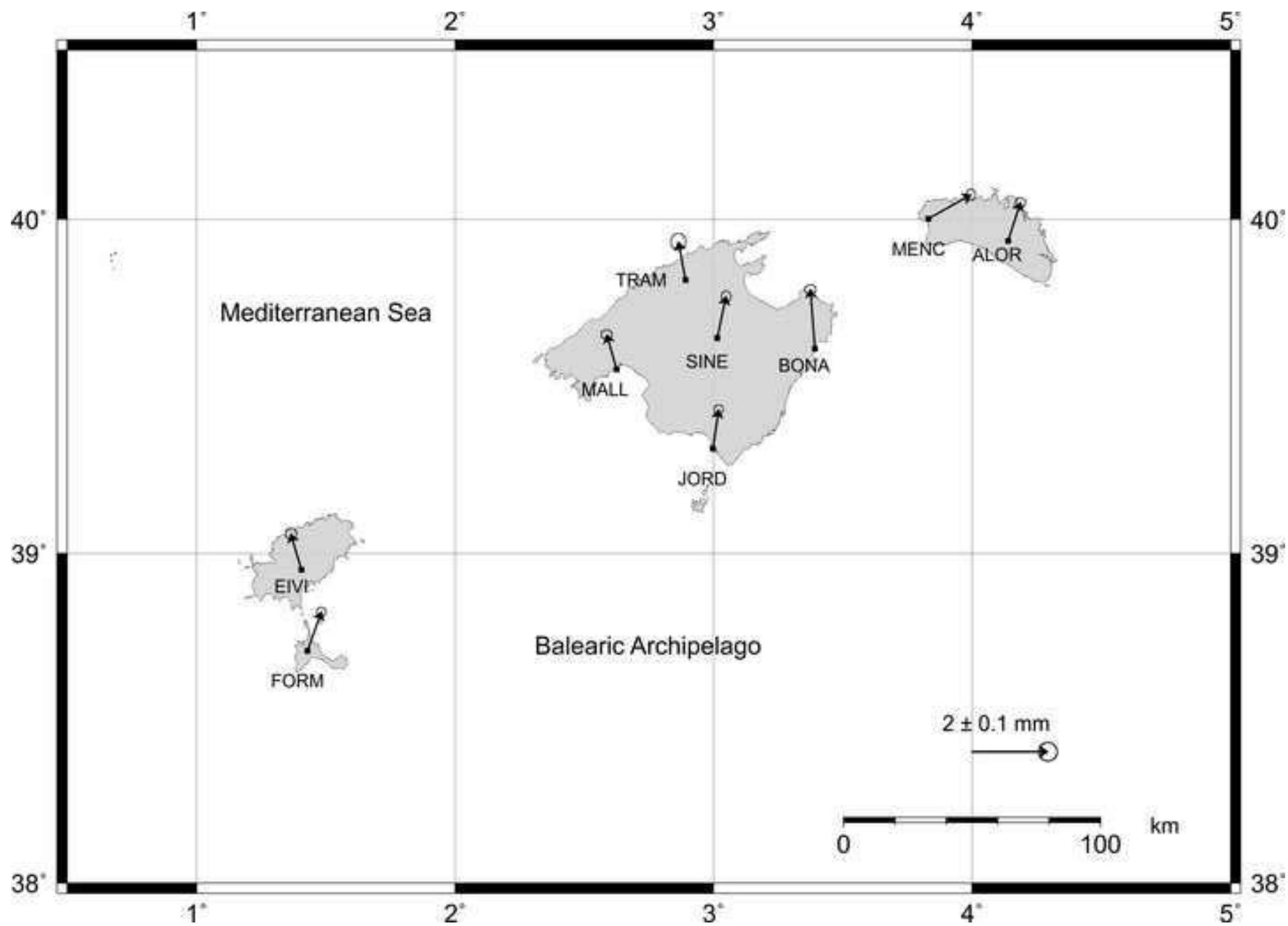

**Figure**

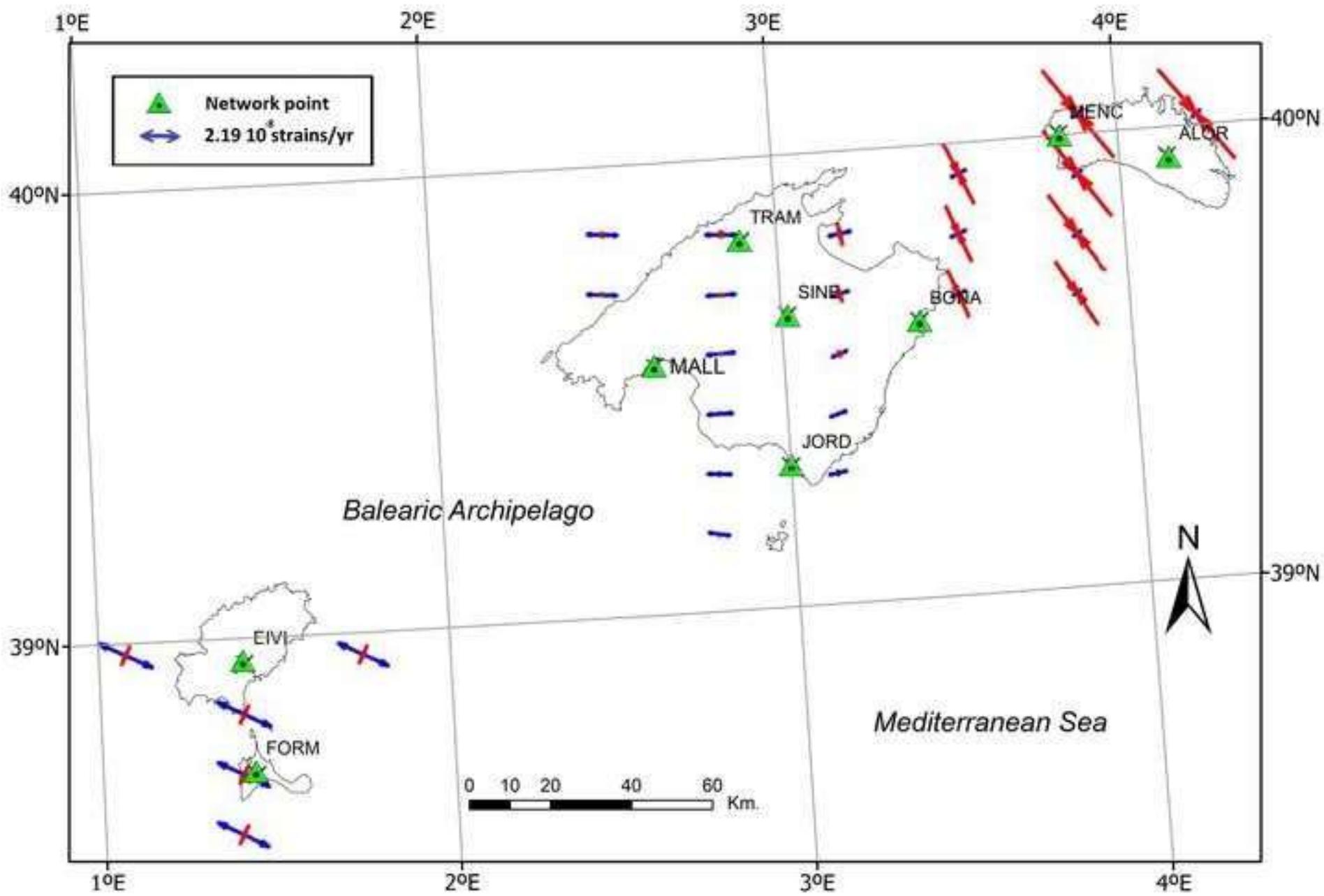

**Figure**

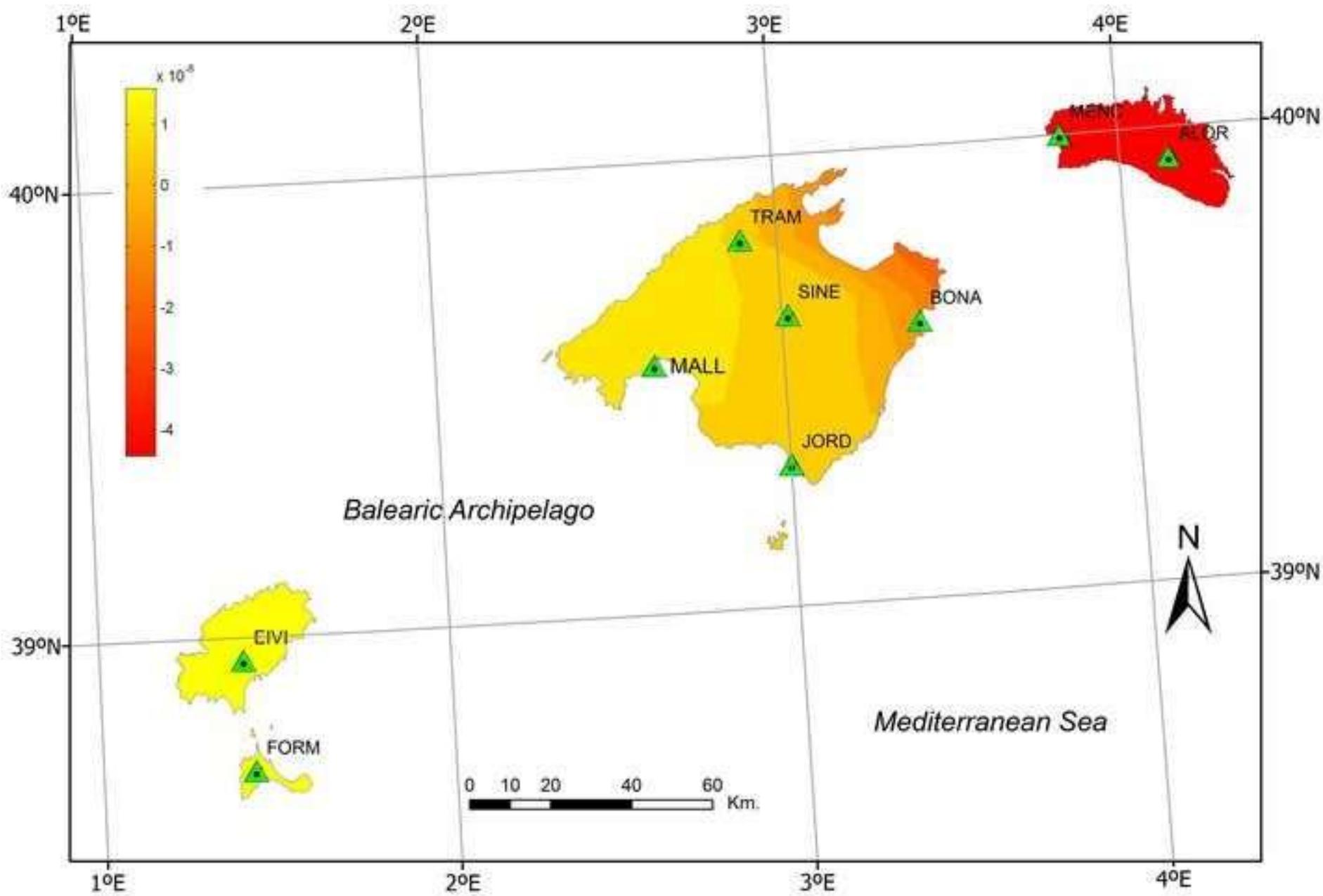

**Figure**

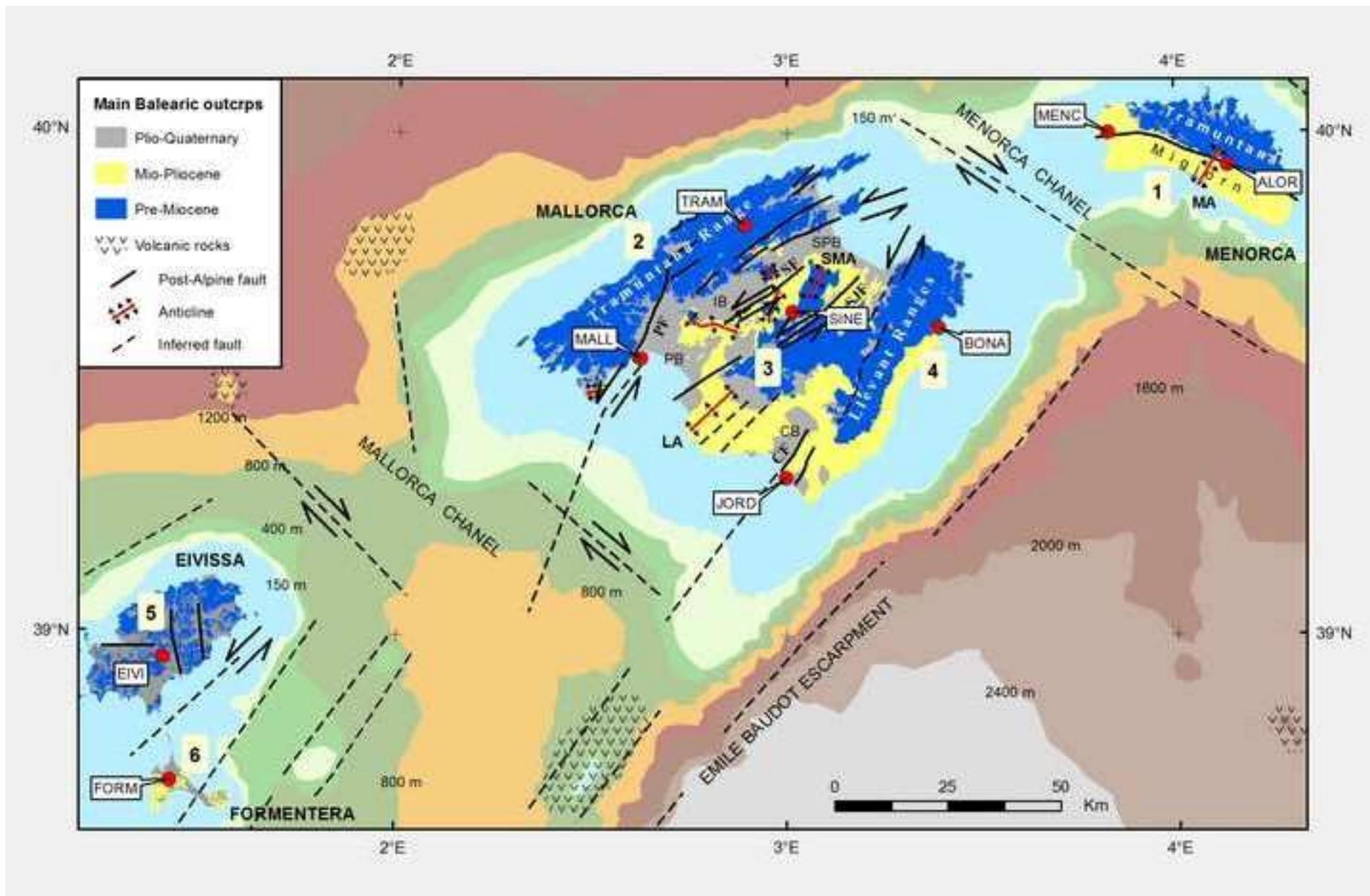